\documentclass[journal]{IEEEtran}

\ifCLASSINFOpdf
\else
   \usepackage[dvips]{graphicx}
\fi
\usepackage{url}
\usepackage{caption}
\usepackage{graphicx}
\usepackage{cite}
\usepackage{amsmath,amssymb,amsfonts}
\usepackage{graphicx}
\usepackage{textcomp}
\usepackage{xcolor}
\usepackage[linesnumbered,ruled,vlined]{algorithm2e}
\usepackage{mathtools}
\DeclarePairedDelimiter\ceil{\lceil}{\rceil}

\begin{document}

\title{Cost-Efficient Deployment of a Reliable Multi-UAV Unmanned Aerial System\\}
\author{Nithin Babu, \IEEEmembership{Student Member, IEEE}, Petar Popovski,  \IEEEmembership{Fellow, IEEE},  Constantinos B. Papadias, \IEEEmembership{Fellow, IEEE}
\thanks{Copyright (c) 2015 IEEE. Personal use of this material is permitted. However, permission to use this material for any other purposes must be obtained from the IEEE by sending a request to pubs-permissions@ieee.org. This version of the work has been accepted for publication in VTC2022-Fall, Workshops. This work is supported by the project PAINLESS which has received funding from the European Union’s Horizon 2020 research and innovation programme under grant agreement No 812991. }
\thanks{N. Babu and C. B. Papadias are with Research, Technology and Innovation Network (RTIN), 
The American College of Greece, Greece (e-mail: nbabu@acg.edu, cpapadias@acg.edu).}
\thanks{N. Babu, C. B. Papadias and P. Popovski are with Department of Electronic Systems, Aalborg University, Denmark (e-mail: niba@es.aau.dk,cop@es.aau.dk,petarp@es.aau.dk)}}
\maketitle
\begin{abstract}
In this work, we study the trade-off between the reliability and the investment cost of an unmanned aerial system (UAS) consisting of a set of unmanned aerial vehicles (UAVs) carrying radio access nodes, called portable access points (PAPs)), deployed to serve a set of ground nodes (GNs). Using the proposed algorithm, a given geographical region is equivalently represented as a set of circular regions, where each circle represents the coverage region of a PAP. Then, the steady-state availability of the UAS is analytically derived by modelling it as a continuous-time birth-death Markov decision process (MDP). Numerical evaluations show that the investment cost to guarantee a given steady-state availability to a set of GNs can be reduced by considering the traffic demand and distribution of GNs.
\end{abstract}
\IEEEpeerreviewmaketitle
\section{Introduction}
Unmanned aerial vehicles (UAVs) carrying radio access nodes, hereafter referred to as portable access points (PAPs), deployed to provide temporary cellular service in remote areas or to assist in emergencies has gained large attention lately \cite{survey1}. The architecture and Quality-of-Service (QoS) requirements for such a system have been proposed in the 3GPP item \cite{3gpp}. Regardless of the limited onboard available energy constraint, the mobile feature of a PAP may enhance the received signal-to-noise ratio (SNR) values through better communication channels compared to a conventional fixed infrastructure system. Since the received SNR values are functions of aerial locations of the PAPs, efficient PAP deployment planning is of paramount importance. In \cite{survey1}, the authors summarize the works that have considered UAV(s) placement problems from an energy efficiency perspective, whereas, \cite{survey2} outlines the works that position UAV(s) to maximize communication-related parameters such as coverage area and throughput. \cite{plos} proposes a general probabilistic line-of-sight (LoS)-non-LoS air to ground channel model and determines the optimal altitude that maximizes the coverage region. The authors of \cite{joint} propose a graph-based algorithm to improve the throughput by jointly optimizing the user association, UAV altitude, and transmission direction, whereas in \cite{babu2} we propose an energy-efficient 3D deployment of a multi-PAP system. From \cite{survey1} - \cite{babu2}, the existing works maximize the respective performance metrics by considering a constant traffic demand. 
However, in practice, the traffic demands from the users are not constant and 
this varying nature of the traffic demand can be efficiently utilized by an unmanned aerial system (UAS) to reduce the investment cost: if the rate at which a new service request arrives is much lower compared to the time required by a PAP to complete the current request (low traffic intensity), then instead of deploying an additional PAP, the mobile feature of the PAP can be used to assign the same PAP to serve the new request after serving the current request thereby reducing the investment cost. 
Hence the main challenge here is to find the optimal number of PAPs required to ensure the availability of at least one idle PAP when a new service request generates represented as a reliability metric called the steady-state availability of the system: the higher the number of PAPs, the higher the steady-state availability. In this work, we determine the minimum investment cost, a function of the number of PAPs, required to guarantee a given availability threshold in different traffic intensity scenarios by exploiting the trade-off between the number of PAPs and the system reliability which has, to the best of our knowledge, not been considered in the literature.

 Section \ref{systemmodel} explains the system model and definitions. 
 In Section \ref{LCP}, firstly, the given geographical region is represented as a set of minimum number of circular cells each representing the coverage region of a PAP using the circle packing theory; then in Section \ref{reliability} a reliability analysis of the UAS is performed to determine the optimal number of PAPs required to guarantee a given steady state availability threshold by modeling the system as a birth-death Markov process. All our main findings from the numerical evaluations are discussed in Section \ref{result}.
 \begin{figure}[t]
\setlength{\belowcaptionskip}{-15pt}
\centering
\captionsetup{justification=centering}
\centerline{\includegraphics[width=0.9\columnwidth]{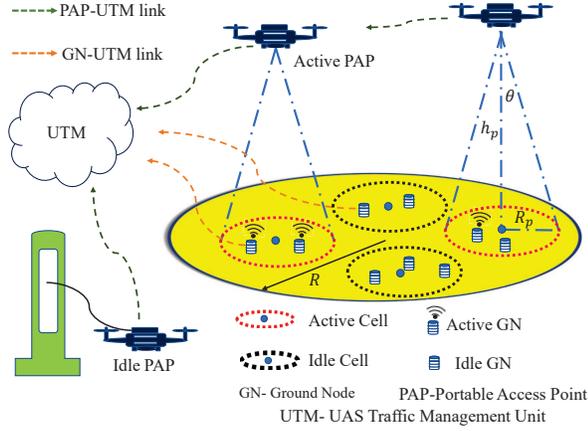}}
\caption{System setup.}
\label{figure1}
\end{figure}
\section{System Model}\label{systemmodel}
A set of $N$ stationary ground nodes (GNs) that are uniformly distributed along a circular geographical region of radius $R$ are considered to be served by a set of PAPs. As shown in Fig. \ref{figure1}, the known locations of all the GNs and the PAPs are registered to a centralised controller unit called the UAS traffic management (UTM) unit \cite{3gpp}. The deployment and positioning of the PAPs according to the traffic demand from the GNs are controlled by the UTM through the PAP-UTM links. Moreover, the PAPs are assumed to be equipped with a directional antenna of half-power beamwidth $2\theta$ with antenna gain in direction $(\psi,\omega)$ given by,
\begin{IEEEeqnarray}{rCl}
G_{\mathrm{p}}& = & \left\{\begin{matrix}
\dfrac{G_{\mathrm{o}}}{\theta^{2}}&-\theta\leq\psi\leq\theta, -\theta\leq\omega\leq\theta,\\
\approx 0 &\text{otherwise},
\end{matrix}\right.
\label{antenna Gain}
\end{IEEEeqnarray}
where $G_{\mathrm{o}}\approx 2.2846$ \cite{joint}. All the GNs are equipped with an omnidirectional antenna and are allocated with orthogonal channels. Consequently, the given geographical area is divided into cells of circular shape representing the possible coverage regions of the PAPs. A cell in which at least one GN is active is called an \textit{active cell}, whereas, a cell in which no GN requires a PAP service is called an \textit{idle cell}. When a GN from a previously idle cell requires the service of a PAP, the request is sent to the UTM through the corresponding GN-UTM link and the UTM sends an activation command to an idle PAP through the PAP-UTM link. The radius (vertical coordinate of a PAP) of the coverage region is jointly determined by the uplink and downlink quality-of-service (QoS) constraints and the horizontal plane coordinates of the PAPs are assumed to be those of the center of the cells.
\subsubsection{Channel model}
 The path loss between a GN located at a distance of $r_{i}$ from the center of a cell and the corresponding PAP hovering at an altitude $h_{\mathrm{p}}$ is taken as the probabilistic mean of the line-of-sight (LoS) and  non-LoS (N-LoS) path loss components \cite{babu2}:
\begin{IEEEeqnarray}{rCl}\label{mpl}
\overline{L}(r_{i},h_{\mathrm{p}}) & = & {P_{\mathrm{l}}\times L_{\mathrm{l}}+\left(1-P_{\mathrm{l}}\right) \times L_{\text{nl}}},\nonumber\\
& = & \underbrace{\frac{r_{i}^{2}+h_{\mathrm{p}}^{2}} {g_{0}}}_{\text{FSPL}} \times \underbrace{[\eta^2_{\text{nl}}+P_{\mathrm{l}}(\eta^2_{\mathrm{l}}-\eta^2_{\text{nl}})],}_{\eta_{\mathrm{m}}(\phi_i),\,\,\text{mean\,\,additional\,\, path\,\, loss}}
\end{IEEEeqnarray}
 where $g_{0}$ represents the channel gain at a reference distance of 1m; $P_{\mathrm{l}}  =  1/\{1+a \exp\left[-b(\phi_{i}-a))\right]\}$ is the LoS probability with $a$ and $b$  being the environment-dependent parameters given in \cite{plos} and $\phi_{i}=(180/\pi)\text{tan}^{-1}(h_{\mathrm{p}}/{r_{i}})$; $\eta_{\mathrm{l}}$ and $\eta_{\text{nl}}$ are the mean value of the additional path loss due to shadowing for LoS and N-LoS links, respectively.
\subsubsection{PAP coverage region} A GN is considered to be in the coverage region of a PAP if it has minimum downlink and uplink SNR values. The SNR received at the $i^{\text{th}}$ GN from a PAP (downlink) is $\frac{G_{\mathrm{p}}P}{\overline{L}(r_{i},h_{\mathrm{p}})\sigma^2}$,
 where $P$ and $\sigma^2$ are the power allocated to a GN and noise power, respectively. From \eqref{mpl}, $\overline{L}(R^{\mathrm{d}}_\mathrm{p},h_{\mathrm{p}})\geq \overline{L}(r_{j},h_{\mathrm{p}})\, \forall r_{j}\leq R^{\mathrm{d}}_\mathrm{p}$. Hence,
for a given downlink QoS constraint, $\Gamma^\mathrm{d}$, considering the SNR at an edge GN, the downlink coverage radius is determined as,
\begin{IEEEeqnarray}{rCl}
\label{R}
R_{\mathrm{p}}^{\mathrm{d}}(\Gamma^{\mathrm{d}},\theta) &=& \sqrt{\frac{G_{\mathrm{p}}g_0 P\text{sin}^{2}\theta}{\Gamma^\mathrm{d} \sigma^2\eta_{\mathrm{m}}(\theta)}}.
\end{IEEEeqnarray}

For the uplink, each GN chooses its transmit power according to the uplink power control specified in the 3GPP technical report \cite{3gpplte}. Then the transmit power for the $i^{th}$ GN (in Watts) is given by, 
 \begin{IEEEeqnarray}{rCl}
\overline{P}_{i} &=&  {P_{a}\overline{L}(r_{i},h_{\mathrm{p}})},
\label{api}
\end{IEEEeqnarray}
where $P_{\mathrm{a}}$ is the target arrived power at the PAP. Accordingly, the received SNR from all the GNs at a PAP will be the same: $(G_{\mathrm{p}}P_{\mathrm{a}})/\sigma^2$. If $\Gamma^{\mathrm{u}}$ is the minimum uplink QoS threshold, then $P_{\mathrm{a}} =\left(\Gamma^{\mathrm{u}}\sigma^2\right)/G_{\mathrm{p}}$. Also, the average power transmitted by a GN should be less than the maximum power $P_{\text{max}}$ that decides the uplink coverage radius $R^{\mathrm{u}}_{\mathrm{p}}$:
\begin{IEEEeqnarray}{L}
  P_{\mathrm{a}}\overline{L}(R_{\mathrm{p}}^{\mathrm{u}},h_{\mathrm{p}}) \leq P_{\text{max}},\nonumber\\
  R_{\mathrm{p}}^{\mathrm{u}}(\Gamma^{\mathrm{u}},\theta) =\sqrt{ \frac{G_{\mathrm{p}}g_0 P_\text{max}\text{sin}^{2}\theta}{\Gamma^{\mathrm{u}}\sigma^2\eta_{\mathrm{m}}(\theta)}}.
\end{IEEEeqnarray}
Therefore the coverage radius of a PAP is given by, $R_{\mathrm{p}} (\Gamma^{\mathrm{d}}, \Gamma^{\mathrm{u}},\theta) = \text{min}\left\lbrace  R_{\mathrm{p}}^{\mathrm{d}}(\Gamma^{\mathrm{d}},\theta),  R_{\mathrm{p}}^{\mathrm{u}}(\Gamma^{\mathrm{u}},\theta)\right\rbrace$ and the corresponding hovering height of the PAP is $h_{\mathrm{p}}=R_{\mathrm{p}} (\Gamma^{\mathrm{d}}, \Gamma^{\mathrm{u}},\theta)\text{tan}\theta$.
\section{Cost Vs Reliability}\label{LCP}
In this section, we analyze the trade-off between the investment cost and reliability of the UAS. The investment cost is determined by the total number of PAPs required to guarantee a given reliability threshold to the GNs. 
 The given circular geographical region needs to be covered by a set of PAP coverage regions of radius $R_{\mathrm{p}} (\Gamma^{\mathrm{d}}, \Gamma^{\mathrm{u}},\theta)$. If $R > R_{\mathrm{p}} (\Gamma^{\mathrm{d}}, \Gamma^{\mathrm{u}},\theta)$, we need multiple cells to cover the given region. 
 With the determined value of $R_{\mathrm{p}} (\Gamma^{\mathrm{d}}, \Gamma^{\mathrm{u}},\theta)$ from Section \ref{systemmodel}, the problem of finding the minimum number of required cells and their locations is solved using a regular-hexagon-based 7-circle multi-level circle packing as detailed in Algorithm 1: the maximum radius of a circle which can be covered by 7 equi-radius circles of radius $R_{\mathrm{p}} (\Gamma^{\mathrm{d}}, \Gamma^{\mathrm{u}},\theta)$ is $R=2R_{\mathrm{p}} (\Gamma^{\mathrm{d}}, \Gamma^{\mathrm{u}},\theta)$; one of these 7 smaller circles will be concentric with the given circular region of radius $R$, and the centers of the remaining 6 circles lie on the vertices of a regular hexagon of side length $\sqrt{3} R_{\mathrm{p}} (\Gamma^{\mathrm{d}}, \Gamma^{\mathrm{u}},\theta)$ \cite{circle}. If $(x_{l-1},y_{l-1})$ is the center of the region to be covered, then centers of the 7 smaller circles of radius $r_{l}$ that covers the given region have the coordinates $\{(x_{j},y_{j})\}$ where, 
\begin{IEEEeqnarray}{L}
x_{j}= x_{l-1}+ r_{l}\sqrt{3}\text{cos}\left(\dfrac{2\pi j}{6}\right) \quad \forall j\in \left\lbrace0,1,...5\right\rbrace,\label{c1}\\
y_{j}= y_{l-1}+r_{l}\sqrt{3}\text{sin}\left(\dfrac{2\pi j}{6}\right)\quad \forall j\in \left\lbrace0,1,...5\right\rbrace, \\
(x_{6},y_{6})=(x_{l-1},y_{l-1}).\label{c3}
\end{IEEEeqnarray}

\begin{figure}
\centering
\captionsetup{justification=centering}
\centerline{\includegraphics[width=0.7\columnwidth]{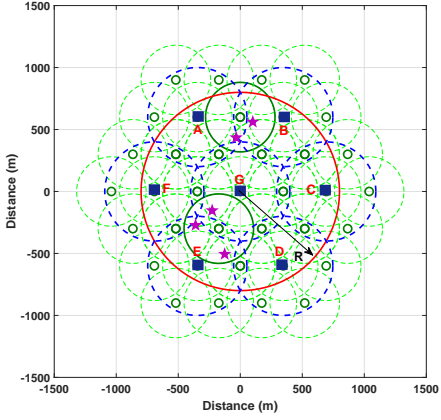}}
\caption{Position of cells for R=800m.}
\label{mcp}
\end{figure}

Algorithm 1 arranges the 7-circle packing pattern on multiple levels as shown in Fig. \ref{mcp}: in the first level, the centers of 7 smaller circles of radius $R/2$ are determined using step 4 (set of blue circles); if the radius of the smaller circle is less than or equal to the coverage radius of a PAP, the packing stops otherwise the packing continues to cover each smaller circle obtained from the previous level (set of green circles). This continues until the radius of the circle to be covered is less than or equal to $R_{\mathrm{p}} (\Gamma^{\mathrm{d}}, \Gamma^{\mathrm{u}},\theta)$. In step 10, the circles representing the cells without any uncovered GNs (dotted green circles) are discarded to obtain the minimum number of cells.

 \underline{\textit{Remark} 1:} The number of levels required to cover a given geographical region of radius $R$ using Algorithm 1 is,\\ $l_{\text{max}}=\ceil*{\text{log}_{2}\left(\dfrac{R}{R_{\mathrm{p}} (\Gamma^{\mathrm{d}}, \Gamma^{\mathrm{u}},\theta)}\right)}.$
 
It should be noted that, with given $R$ and $\Gamma^{\mathrm{d}}, \Gamma^{\mathrm{u}}$, and $\theta$, step 2 to 9 takes a running time of $O\left(l_{\text{max}}\right)$, whereas step 10 requires $\sum\limits_{c=0}^{7^{l_{\text{max}}}-1}(7^{l_{\text{max}}}-c)(7^{l_{\text{max}}}-c-1)$ pair checks; hence the complexity of Algorithm 1 is, at worst 
$O\left[l_{\text{max}}+7^{2l_{\text{max}}}\right]$.
\begin{algorithm}[]
\caption{Optimal Number of Cells}
\textbf{Input}: $R$, $R_{\mathrm{p}} (\Gamma^{\mathrm{d}}, \Gamma^{\mathrm{u}},\theta), l=0$, $\mathcal{C}_{l}=\{(0,0)\}$; \\
\While{1}
{
$l=l+1;$ $r_{l}=R/2$;\\
For each $\mathbf{c}_{l-1}=(x_{l-1},y_{l-1})\in \mathcal{C}_{l-1}$, find the set of center of PAP coverage circles using \eqref{c1} - \eqref{c3}: $\mathcal{C}_{l}$;\\
\If{$r_{l} \leq R_{\mathrm{p}} (\Gamma^{\mathrm{d}}, \Gamma^{\mathrm{u}},\theta)$}
{
break;
}
\Else
{
$R=R/2$; \\
goto Step 3;\\
}
}
Remove the circles with zero non-shared GNs;\\
\textbf{Output}:{Total number of cells, $n=|\mathcal{C}_{l}|$}.
\end{algorithm} 
\subsection{Reliability Analysis and Optimal Number of PAPs.}\label{reliability}
\begin{figure}
\centering
\captionsetup{justification=centering}
\centerline{\includegraphics[width=0.9\columnwidth]{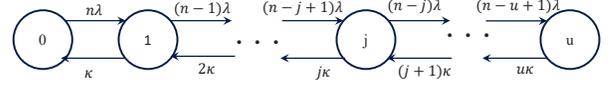}}
\caption{State transition diagram of the UAS.}
\label{figure2}
\end{figure}

In this section, we find the optimal number of PAPs required to guarantee a given reliability threshold to the GNs distributed along $n$ cells. 
Let $A(t)$ be an event that a PAP is available for serving the request from a new active cell at time $t$; hence, the steady-state availability is defined as, 
    \begin{IEEEeqnarray}{rCl}
       A &=& \lim_{t\to\infty} \mathcal{P}[A(t)],
    \end{IEEEeqnarray}
 where $\mathcal{P}[A(t)]$ is the probability of $A(t)$; here, the unavailability ($1-A$) is only due to busy PAPs, but not the wireless environment. We consider $n$ cells determined using Algorithm 1 as $n$ sources that generate $n$ independent Poisson inputs with common intensity $\lambda$. Also, the service times of PAPs at cells are considered to be independent, exponentially distributed random variables with parameter $\kappa$. The deployment of an additional PAP is only required when a new request originates from an idle cell. 
Additionally, the probability that more than one idle cell becomes active or vice versa is assumed to be negligible. Let $u$ be the number of PAPs available in the UAS to serve $n$ cells; 
thus the considered system can be modelled as a birth-death Markov process as shown in Fig. \ref{figure2} with state-space $Z=\{0,1,2,..u\}$ representing the number of serving PAPs and transition rates, 
\begin{IEEEeqnarray}{rCl}
& & \lambda_j=(n-j)\lambda;\quad \forall j=\{0,1,2,...(u-1)\},\\
& & \kappa_j =  j\kappa\quad \forall j=\{1,2,...u\}.
\end{IEEEeqnarray}
Let $\{p_j, j\in Z\}$ be the steady-state probability distribution of the considered Markov process. Then the state equations from Fig. \ref{figure2} are expressed as \cite{markov},
\begin{IEEEeqnarray}{L}
\kappa p_1 = n\lambda p_0, \label{p1} \\\label{p1}
(n-j+1)\lambda p_{j-1} + (j+1)\kappa p_{j+1} = \left(j\kappa +(n-j)\lambda\right) p_{j},\nonumber\\
(n-u+1)\lambda p_{u-1} = u\kappa p_{u}.
\end{IEEEeqnarray}
From \eqref{p1}, $p_{1}=n\delta p_{0}$, where $\delta=\lambda/\kappa$. By successive substitution we get,
\begin{IEEEeqnarray}{rCl}
p_{j} &=& C(n,j)\delta^j p_{0}\quad \forall j \in \mathcal{Z}\label{pj},
\end{IEEEeqnarray}
where $C(n,j)=n!/(j!(n-j)!)$. For a finite $R$ value, the number of cells will be finite; hence, the stationary state probabilities should satisfy the normalizing condition $\Sigma_{j=0}^{u}p_{j}=1$:
\begin{IEEEeqnarray}{L}
p_{0}+\Sigma_{j=1}^{u}C(n,j)\delta^j p_{0}=1,\\
p_{0}=\frac{1}{1+\Sigma_{j=1}^{u}C(n,j)\delta^j}\label{p0}.
\end{IEEEeqnarray}
Let $F$ be an event that an active cell does not find an available PAP, then using \eqref{pj} and \eqref{p0},
\begin{IEEEeqnarray}{rCl}
P(F) & = & p_{u} = \frac{C(n,u)\delta^u}{1+\Sigma_{j=1}^{u}C(n,j)\delta^j}.\label{pf}
\end{IEEEeqnarray}
Hence, the steady-state availability of the system can be determined as, 
\begin{IEEEeqnarray}{rCl}
A & = & 1-P(F) = 1-\frac{C(n,u)\delta^u}{1+\Sigma_{j=1}^{u}C(n,j)\delta^j}.\label{A}
\end{IEEEeqnarray}
Thus, for a given availability threshold, $\rho$, the optimal number of PAPs required is obtained using \eqref{A} as,
\begin{IEEEeqnarray}{rCl}
u_{\text{opt}}&=& \text{min}\{(u|A>=\rho), n\}\label{uopt}.
\end{IEEEeqnarray}
The degree of PAP utilization defined as the ratio of the mean number of deployed PAPs to the total number of PAPs is given by,
\begin{IEEEeqnarray}{rCl}
\eta (u_{\text{opt}})&=& \frac{\Sigma_{j=1}^{u_{\text{opt}}}jp_{j}}{u_{\text{opt}}}=\frac{\Sigma_{j=1}^{u_{\text{opt}}}j C(n,j)\delta^j}{u_{\text{opt}}\left(1+\Sigma_{j=1}^{u_{\text{opt}}}C(n,j)\delta^j\right)}\label{eta}.
\end{IEEEeqnarray}
\section{Numerical Evaluation and Conclusion}\label{result}
\begin{figure}
\centering
\captionsetup{justification=centering}
\centerline{\includegraphics[width=0.9\columnwidth]{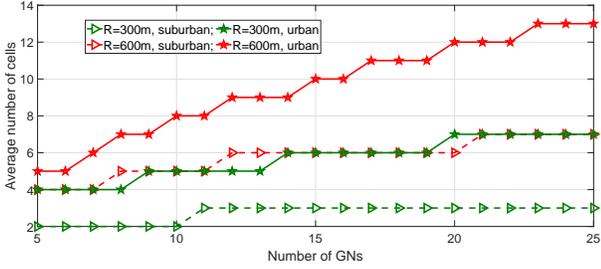}}
\caption{Average number of cells Vs number of GNs .}
\label{resl_fig1}
\end{figure}

In this section, we present our main findings from numerical evaluations.The considered simulation parameters are $g_{0}=1.42\times 10^{-4}$,\,$\Gamma^{\mathrm{d}}=\Gamma^{\mathrm{u}}=100$, $\sigma^2=1.25\times 10^{-14} W$, $P=1$mW, $P_{\text{max}}=1W$. Fig. \ref{resl_fig1} shows the variations of the average number of cells required (rounded to the nearest integer) as a function of the number of GNs for suburban and urban deployment scenarios; the corresponding $(a,b,\eta_{\mathrm{l}},\eta_{\mathrm{nl}},\theta)$ parameters used are $(4.83,0.43,1.01,11.22,70^{o})$, $(9.6,0.16,1.12,10,52^{o})$ \cite{plos}. The averaging is done over the minimum number of cells obtained using Algorithm 1 for 600 independent realizations of GNs. From the figure, for a given number of GNs, the number of cells required increases as the radius of both the suburban and urban region increase because of the widely distributed GNs in a larger geographical region. 
Additionally, for a given QoS threshold, the number of cells required to cover any $N$ GNs in the urban region is higher than that in the suburban region: this is because the radius of the coverage region of a PAP decreases as we move from suburban to urban region due to an increase in the number and density of buildings ($\eta_m(\theta)$).   

\begin{figure}
\centering
\captionsetup{justification=centering}
\centerline{\includegraphics[width=0.9\columnwidth]{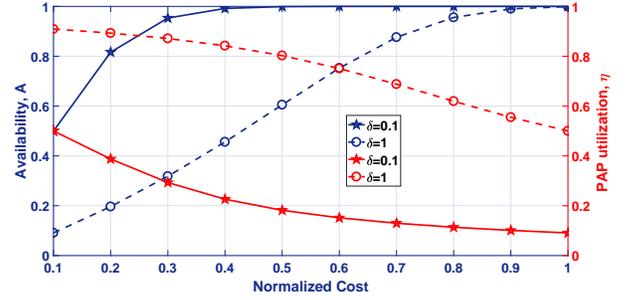}}
\caption{Availability Vs cost reduction.}
\label{resl_fig2}
\end{figure}

Fig. \ref{resl_fig2} is plotted considering 10 cells ($R=600$m, $N=15$, urban) and it shows the variation of the availability and PAP utilization factor, given by \eqref{A} and \eqref{eta}, as a function of the normalized investment cost; the normalized investment cost is taken as the ratio of $u_{\text{opt}}$ to the total number of cells.  
As seen in the figure, the steady-state availability increases with deploying more PAPs (more investment cost). However, for low traffic intensity scenarios, $\delta=0.1$, a steady-state availability of 1 is achievable with PAPs equal to half the number of cells. This benefit of investment cost reduction becomes less with an increase in the traffic intensity: since the rate of arrival of the new service request is higher than the service duration, the probability of finding an idle PAP to serve the request from a new active cell is very small thereby demanding a deployment scheme that assigns one dedicated PAP per cell. 

\begin{figure}
\centering
\captionsetup{justification=centering}
\centerline{\includegraphics[width=0.9\columnwidth]{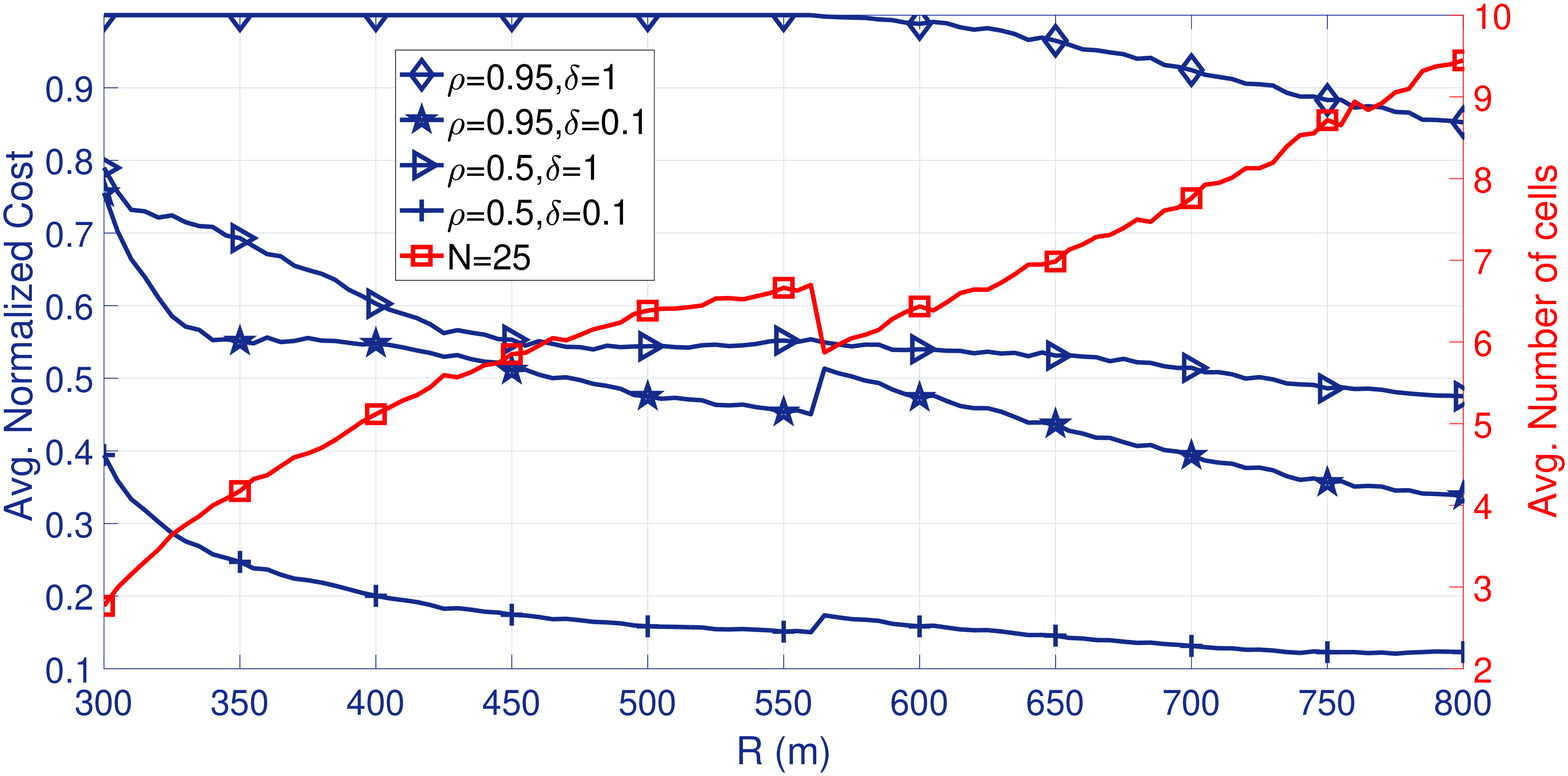}}
\caption{Normalized cost Vs radius of the region.}
\label{resl_fig3}
\end{figure}

Fig. \ref{resl_fig3} is plotted by considering a suburban deployment scenario with 25 GNs; it shows the variation of the average normalized cost incurred to guarantee a given availability threshold to GNs for different traffic intensity scenarios. As expected, the average number of cells required increases with an increase in $R$. Both the availability threshold and the traffic intensity decide the investment cost; the cost can be reduced either if the intensity of the traffic demand from the GNs is low or if the availability threshold is light. Moreover, the cost reduction is more noticeable if the GNs are widely separated. 
Hence, from Fig. \ref{resl_fig2} and Fig. \ref{resl_fig3}, the number of PAPs required should be economically determined by considering not only the area of the geographical region but also the traffic intensity and the distribution of the GNs in the considered region. 

For a given geographical region with a set of GNs, the minimum number of cells required is determined using Algorithm 1, and depending on the type of service required by the GNs and the traffic intensity, a suitable availability threshold could be selected which in turn decides the optimal number of required PAPs and the corresponding investment cost. The optimal number of PAP determination by considering a general distribution of the traffic intensity is left as future work.  

\bibliographystyle{IEEEtran}
\bibliography{./main.bbl}

\begin{thebibliography}{1}
\providecommand{\url}[1]{#1}
\csname url@samestyle\endcsname
\providecommand{\newblock}{\relax}
\providecommand{\bibinfo}[2]{#2}
\providecommand{\BIBentrySTDinterwordspacing}{\spaceskip=0pt\relax}
\providecommand{\BIBentryALTinterwordstretchfactor}{4}
\providecommand{\BIBentryALTinterwordspacing}{\spaceskip=\fontdimen2\font plus
\BIBentryALTinterwordstretchfactor\fontdimen3\font minus
  \fontdimen4\font\relax}
\providecommand{\BIBforeignlanguage}[2]{{%
\expandafter\ifx\csname l@#1\endcsname\relax
\typeout{** WARNING: IEEEtran.bst: No hyphenation pattern has been}%
\typeout{** loaded for the language `#1'. Using the pattern for}%
\typeout{** the default language instead.}%
\else
\language=\csname l@#1\endcsname
\fi
#2}}
\providecommand{\BIBdecl}{\relax}
\BIBdecl

\bibitem{survey1}
S.~Shakoor, Z.~Kaleem, M.~I. Baig, O.~Chughtai, T.~Q. Duong, and L.~D. Nguyen,
  ``{Role of UAVs in public safety communications: Energy efficiency
  perspective},'' \emph{IEEE Access}, vol.~7, pp. 140\,665--140\,679, 2019.

\bibitem{3gpp}
``{3GPP; Technical Specification Group Services and System Aspects}; unmanned
  aerial system (uas) support,'' \emph{Stage 1; Release 17}, 2017.

\bibitem{survey2}
C.~T. Cicek, H.~Gultekin, B.~Tavli, and H.~Yanikomeroglu, ``{UAV Base Station
  Location Optimization for Next Generation Wireless Networks: Overview and
  Future Research Directions},'' in \emph{2019 1st International Conference on
  Unmanned Vehicle Systems-Oman (UVS)}.\hskip 1em plus 0.5em minus 0.4em\relax
  IEEE, 2019, pp. 1--6.

\bibitem{plos}
A.~Al-Hourani, S.~Kandeepan, and S.~Lardner, ``{Optimal LAP Altitude for
  Maximum Coverage},'' \emph{IEEE Wireless Communications Letters}, vol.~3,
  no.~6, pp. 569--572, 2014.

\bibitem{joint}
W.~Huang, D.~M. Kim, W.~Ding, and P.~Popovski, ``{Joint Optimization of
  Altitude and Transmission Direction in UAV-Based Two-Way Communication},''
  \emph{IEEE Wireless Communications Letters}, vol.~8, no.~4, pp. 984--987,
  2019.

\bibitem{babu2}
N.~Babu, C.~B. Papadias, and P.~Popovski, ``{Energy-Efficient 3D Deployment of
  Aerial Access Points in a UAV Communication System},'' \emph{IEEE
  Communications Letters}, vol.~24, no.~12, pp. 2883--2887, 2020.

\bibitem{3gpplte}
``{3GPP, Physical Layer Procedures},'' \emph{TR 36.213, Sep. 2015, v 10.12},
  2017.

\bibitem{circle}
G.~F. T{\'o}th, ``{Thinnest Covering of a Circle by Eight, Nine, or Ten
  Congruent Circles},'' \emph{Combinatorial and computational geometry},
  vol.~52, no. 361, p.~59, 2005.

\bibitem{markov}
F.~Beichelt, \emph{{Stochastic Processes in Science, Engineering and
  Finance}}.\hskip 1em plus 0.5em minus 0.4em\relax Chapman and Hall/CRC, 2006.

\end{thebibliography}
\end{document}